\documentclass[
]{ceurart}

\begin{document}

\copyrightyear{2021}
\copyrightclause{Copyright for this paper by its authors.
  Use permitted under Creative Commons License Attribution 4.0
  International (CC BY 4.0).}

\conference{LWDA'21: Lernen, Wissen, Daten, Analysen
  September 01--03, 2021, Munich, Germany}

\title{Exploiting Sentence-Level Representations for Passage Ranking}

\author[1]{Jurek Leonhardt}[%
email=leonhardt@l3s.de,
]
\author[1]{Fabian Beringer}[%
email=beringer@l3s.de,
]
\author[1]{Avishek Anand}[%
email=anand@l3s.de,
]
\address[1]{L3S Research Center, Appelstraße 9a, 30167 Hannover, Germany}

\begin{abstract}
    Recently, pre-trained contextual models, such as BERT, have shown to perform well in language related tasks. We revisit the design decisions that govern the applicability of these models for the \emph{passage re-ranking} task in open-domain question answering. We find that common approaches in the literature rely on fine-tuning a pre-trained BERT model and using a single, global representation of the input, discarding useful fine-grained relevance signals in token- or sentence-level representations. We argue that these discarded tokens hold useful information that can be leveraged. In this paper, we explicitly model the sentence-level representations by using Dynamic Memory Networks (DMNs) and conduct empirical evaluation to show improvements in passage re-ranking over fine-tuned vanilla BERT models by memory-enhanced explicit sentence modelling on a diverse set of open-domain QA datasets. We further show that freezing the BERT model and only training the DMN layer still comes close to the original performance, while improving training efficiency drastically. This indicates that the usual fine-tuning step mostly helps to aggregate the inherent information in a single output token, as opposed to adapting the whole model to the new task, and only achieves rather small gains.
\end{abstract}


\newcommand{\bert}{$\text{BERT}$}
\newcommand{\blite}{$\text{BERT}_\text{lite}$}
\newcommand{\bdmn}{$\text{BERT-DMN}$}
\newcommand{\bdlite}{$\text{BERT-DMN}_\text{lite}$}

\maketitle
\section{Introduction}
Language model pre-training has attracted wide attention and fine-tuning on pre-trained language model has shown to be effective for improving many downstream natural language processing tasks. BERT \cite{devlin2018bert} obtained new state-of-the-art results on a broad spectrum of diverse tasks, offering pre-trained deep bidirectional representations which are conditioned on both left and right context in all layers, which is often followed by discriminative fine-tuning on each specific task, including passage re-ranking for open domain QA.

There are two limitations of using fine-tuned BERT models for re-ranking passages in QA. Firstly, passages are of variable lengths, which affects the quality of BERT-based representations. Specifically, in the fine-tuning regime of BERT for open domain QA and passage re-ranking, a representation is learnt for the entire passage given a question. While this is desirable for small passages or questions that have short and easy answers, it isn't for instances where the passage answers a question using multiple, more complex statements. Secondly, the passage re-ranking task is unlike other QA tasks, like factoid QA and reading comprehension, in that the answers are not limited to a word, phrase or sentence. Potential answers can have varying granularity and passages are judged by annotators based on the likelihood of containing the relevant answer. Therefore, the applicability of vanilla BERT models to answering queries that span multiple sentences or might need reasoning across distant sentences in the same passage is limited.

In this paper we deal with the above problems by extending the BERT model to explicitly model sentence representations. This is realized by distilling the sentence representations from the output of the BERT block and aggregating the representations of the tokens that make a sentence. Secondly, once we have the sentence representations, we apply a \textit{Dynamic Memory Network} \cite{kumar2016ask,xiong2016dynamic} to model sentence-wise relations for relevance estimation. We are interested in the following research questions:
\begin{itemize}
    \item By aggregating BERT representations on a sentence level and then reasoning over sentence representations, can we improve re-ranking performance?
    \item Can we improve training efficiency by light-weight reasoning instead of fine-tuning all parameters of BERT?
\end{itemize}

We perform experimentation on three diverse open-domain QA datasets and show that the sentence-level representations improve the model's re-ranking performance. We find that explicit sentence modeling using a DMN enables us to reason about the answers that spread across sentences. Additionally, we find that \bdmn{}, although being an extension of BERT, can be used without expensive fine-tuning of the BERT model, resulting in reduced training times. The code will be made publicly available.

 \section{Related Work}
 \label{sec:rel-work}
Recent practices in open-domain question answering (QA) can be traced to the Text Retrieval Conferences (TRECs) in the late 1990s. \citet{Voorhees99thetrec-8} defines the task of textual open-domain question answering as using a small text snippet, usually an excerpt from a document as part of a large collection that is being utilized in the process, such as web pages \cite{kwok2001scaling}. In the last decade, the focus on open-domain question answering has shifted to the re-ranking stage, where answer identification from candidate documents is performed using learning strategies based on richer and better language understanding models \cite{tan2016improved,tran2018multihop,wang2018evidence,wang2018r,lin2018denoising}. Our approach also tries to propose models that improve the re-ranking part of the QA pipeline. Specifically, we are different from alternate approaches that perform \textit{end-to-end} question answering that requires some type of term-based retrieval technique to restrict the input text under consideration \cite{chen2017reading,wang2017joint,kratzwald-feuerriegel-2018-adaptive}.

Multiple approaches have been proposed to improve re-ranking in open-domain QA. In \cite{tan2016improved}, the authors use LSTMs to encode questions and answers and then perform attention- and CNN-based pooling in order to perform question-answer-matching; \cite{tran2018multihop} follows a similar idea, but produces multiple vector representations for each question and answer, which can then focus on different aspects. Other works like \cite{lin2018denoising} aim to improve the answer selection process by filtering out noisy, irrelevant paragraphs. Afterwards, the answer is selected from the remaining, relevant paragraphs. Some works have used evidence aggregation to re-rank passages based on information from multiple other passages \cite{wang2018evidence} or reinforcement learning to jointly train a \textit{ranking model} to rank the passages and a \textit{reading model} to extract the answer from a passage \cite{wang2018r}. In \cite{xu2019passage}, the authors use weak supervision to train a BERT-based passage ranking model without any ground-truth labels.

The most recent improvement in re-ranking stage of open-domain QA comes from BERT models that have been shown to improve language understanding. Recent works have used BERT-based ranking models, dealing with efficiency \cite{guo2020detext} and analyzing the attention mechanism \cite{zhan2020analysis}. In \cite{DBLP:conf/rep4nlp/PetersRS19}, the authors compare the performance of BERT with and without fine-tuning on various NLP tasks. \cite{macavaney2019contextualized:bertir} deals with combining traditional Ranking models with BERT token representations.

Neural architectures for document ranking can be roughly categorized into \textit{representation-based} models for learning semantic representations of the text \cite{Shen2014a,dssm13,Shen2014b}, \textit{interaction-based} models for learning salient interaction patterns from the local interactions between the query and document \cite{KNRM17,Guo2016} or a combination of both \cite{mitra2017learning}. Other works \cite{matchpyramid16,Nie_ictir18,Nie_sigir_2018} try to capture hierarchical matching patterns based on \textit{n-gram} matches from the local interaction matrix of the query-document. More recent approaches \cite{pacrr17,co_pacrr_wsdm18,pacrr_drmm_18} have tried to exploit positional information and context of the query terms. Other approaches include query modeling techniques \cite{diaz16,Zamani_16a} with a query expansion based language model (QLM) that uses word embeddings.

\section{Approach}
\label{sec:approach}
The usual question answering process consists of multiple stages. Given a query, a simple method (like BM25) is used to rank a number of passages with respect to the query. Next, the top-$n$ of these passages are \textbf{re-ranked} using a more expensive model. Finally, the top-$k$ ($k < n$) of the re-ranked passages are used to answer the query. This work deals with the passage re-ranking step.

BERT-based models have achieved high performance in passage re-ranking tasks. We find, however, that these models are limited: Firstly, most variants tend to rely solely on BERT's dedicated classification output, operating under the assumption that its internal capabilities of compressing all query and passage representations into a single output are optimal. Secondly, BERT models are very large, which results in slow training.

In this paper we introduce a re-ranking approach that leverages the representations obtained from BERT and aggregates them using a Dynamic Memory Network. We describe DMNs and outline how they can be combined with BERT such that, in addition to the classification output, the query and passage representations are taken into account. Moreover, we investigate how our model can reduce training time by introducing a \textit{lite} version.

\subsection{Dynamic Memory Networks}
\begin{figure}
	\centering
    \includegraphics[width=0.6\linewidth]{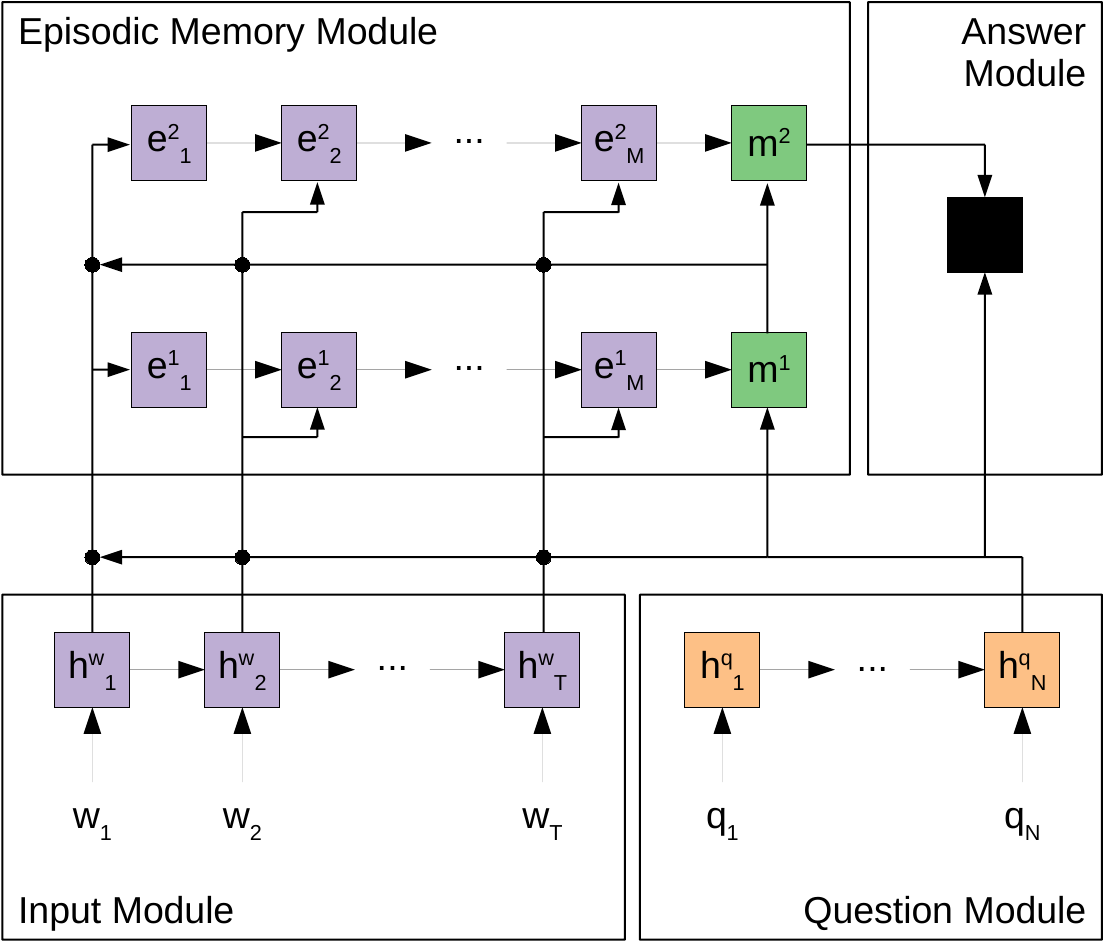}
	\caption{The Dynamic Memory Network architecture with two episodes. The black dots represent the concatenation of vectors. This figure depicts the case where each vector from the input module is used, i.e.\ $T = M$. In the case of a multi-sentence input, $T$ is greater than $M$.}
	\label{fig:dmn}
\end{figure}
In this section we briefly introduce Dynamic Memory Networks \cite{kumar2016ask,xiong2016dynamic}, which we use to aggregate BERT outputs. DMNs take as input a sequence of words $w = (w_1, ..., w_T)$, which usually represent multiple sentences such as a document or a passage, and a question $q = (q_1, ..., q_N)$. They are composed of four \textit{modules} (cf. Figure \ref{fig:dmn}):

The \textbf{input module} encodes the input words as a sequence of vector representations. The input text is represented by pre-trained word embeddings and fed into a word-level many-to-many GRU. The outputs $h^w_t = \mathtt{GRU}(w_t, h^w_{t-1})$ are then used as inputs in other modules. If the input consists of a single sentence, each of the GRU outputs is used; however, if the input consists of multiple sentences, only those GRU outputs $h_t$ are used where $t$ corresponds to an end-of-sentence token (for example periods or question marks), while the rest is discarded. We denote the final sequence of vectors produced by the input module as $s = (s_1, ..., s_M)$.

The \textbf{question module} is similar to the input module, as it is used to encode the query (or question) as a fixed-size vector representation. The word embeddings are fed into a many-to-one GRU, which outputs the query representation $Q$ at the end, i.e.\ $h^q_t = \mathtt{GRU}(q_t, h^q_{t-1})$ and $Q = h^q_N$.

The \textbf{episodic memory module} maintains a number of \textit{episodes}. An episode $e^i$ produces a \textit{memory} $m^i$ by iterating a GRU over the fact representations from the input module, while taking the previous memory $m^{i-1}$ into account. For this, the GRU's update gate is replaced by a special attention gate at each time step,
\begin{equation}
    \mathtt{AttGRU}^i(x_t, h_{t-1}) = g^i_t \circ h'_t + (1 - g^i_t) \circ h_{t-1},
\end{equation}
where $h'_t$ is the candidate hidden state for the GRU's internal update at time step $t$. The \textit{attention gate} $g^i_t$ is a function of the input $x_t$ and the memory and question vectors, encoding their similarities (details can be found in \cite{kumar2016ask}). The initial memory is initialized as $m^0 = Q$. The hidden state of an episode $e^i$ is then computed as $e^i_t = \mathtt{AttGRU}^i(c_t, e_{t-1})$, where $c_t = [s_t; m^{i-1}]$ is a \textit{candidate fact} and $[\cdot; \cdot]$ denotes concatenation. The new memory value is then simply set to the last hidden state of the episode, i.e.\ $m^i = e^i_M$. Finally, the output of the episodic memory module is the last output of a GRU that iterates over all memories $m^i$.

The \textbf{answer module} generates the final output of the model and is therefore highly dependent on the task. In our case, it is a simple feed-forward layer that predicts a score to rank passages given the output of the episodic memory module.

\subsection{Combining BERT and DMN}
\label{sec:bert-dmn}
\begin{figure}
	\centering
     \includegraphics[width=0.6\linewidth]{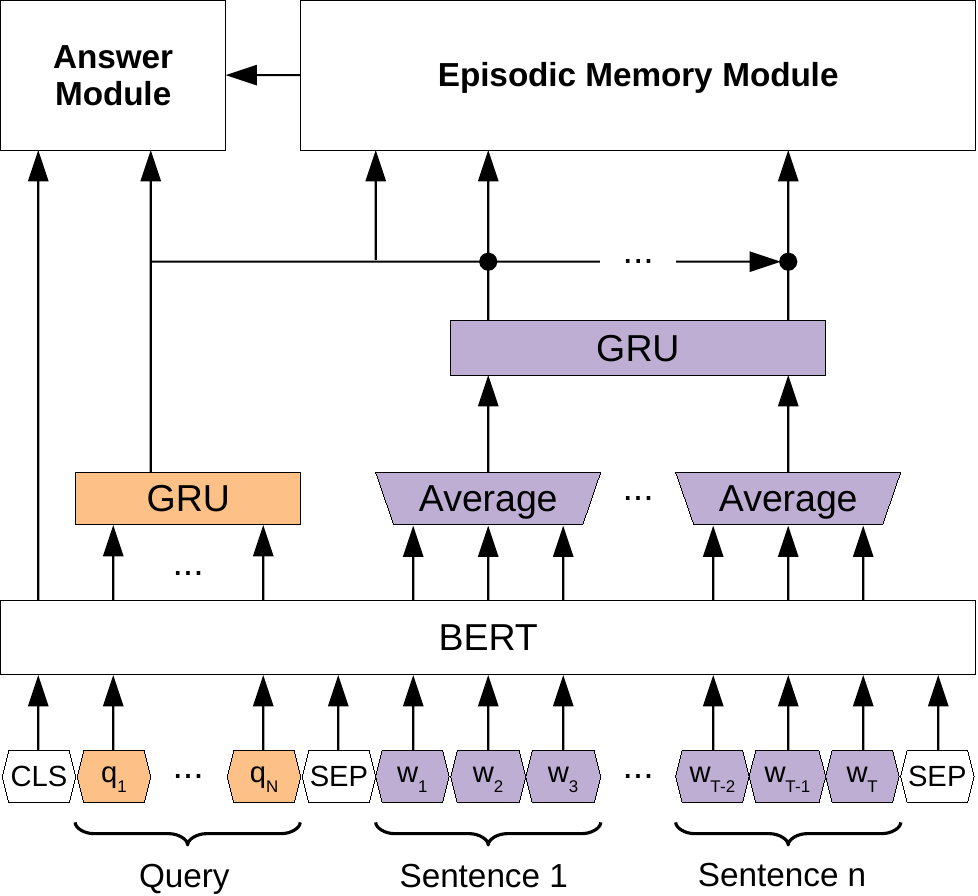}
	\caption{The BERT-DMN model architecture. Note that the padding tokens are omitted here.} 
	\label{fig:model}
\end{figure}
Dynamic Memory Networks have proven to be effective in QA tasks such as reading comprehension. In this paper we combine a DMN with contextualized representations, specifically the outputs of BERT, by modifying the input, question and answer module. The resulting model, \bdmn{}, takes all outputs of BERT into account (including the classification token). It processes the token-level outputs by creating query and sentence representations and reasons over them. In the final step, everything is combined to produce the final query-passage score, which is then used to rank the documents. Figure \ref{fig:model} shows the architecture of our approach.

\subsubsection{Input and Question Module}
Let the query and passage again be denoted by $q = (q_1, ..., q_N)$ and $w = (w_1, ..., w_T)$. We first construct the input for BERT as
\begin{equation}
    \texttt{[CLS]}, q_1, ..., q_N, \texttt{[SEP]}, w_1, ..., w_T, \texttt{[SEP]}.
\end{equation}
Note that $q_i$ and $w_i$ are not necessarily words, as BERT uses subword tokenization. This input format is identical to the usual way BERT is used, where the first input is a classification token, followed by two text inputs, which are separated by separator tokens.

We split the BERT output $o = (o_1, ..., o_L)$ back into two chunks, where one corresponds to the query and the other one to the passage. The outputs corresponding to the \texttt{[SEP]} tokens are discarded. We then use the token representations output by BERT as a replacement for the word embeddings in the DMN. In practice, instead of simply using the vector corresponding to the end-of-sentence token to represent the whole sentence, we take the vectors of all tokens in this sentence and average them.

\subsubsection{Answer Module}
Since the original DMN model was used for reading comprehension tasks, the answer module consisted of a sequence generation network. For the re-ranking task we are only interested in predicting a score, therefore we modify the answer module: Let the final memory be a vector $m \in \mathbb{R}^{d \times 1}$ and the BERT output $c \in \mathbb{R}^{b \times 1}$ correspond to the \texttt{[CLS]} token. We concatenate these vectors along with the query representation $Q$ and compute the final score $a$ using a feed-forward layer, i.e.
\begin{equation}
    a = \sigma(W_a [c; Q; m] + b_a),
\end{equation}
where $\sigma$ is the sigmoid function and $W_a \in \mathbb{R}^{1 \times (b + 2d)}$.

\section{Experimental Setup}
\label{sec:experiments}
In this section we describe the datasets we use for our experiments and the baselines. We further outline the training process.

\subsection{Datasets}
\label{sec:datasets}
\begin{table}
    \centering
    \begin{tabular}{lccc}
        \toprule
                                        & ANTIQUE       & InsuranceQA       & TREC-DL \\
        \midrule
        No. of train queries            & 2406          & 12889             & 808731 \\
        No. of test queries             & 200           & 2000              & 200 \\
        Avg. query length               & 10.55         & 8.42              & 6.53 \\
        \midrule
        No. of passages                 & 33642         & 27413             & 8841823 \\
        Avg. passage length             & 47.83         & 103.59            & 64.63 \\
        \midrule
        Avg. no. of passages per query  & 32.95         & 500               & 1000 \\
        Avg. no. of relevant passages   & 9.6           & 1.66              & 1.69 \\
        \bottomrule
    \end{tabular}
    \caption{Dataset statistics. InsuranceQA and TREC-DL have dedicated devsets; for ANTIQUE, we use a small fraction of the trainset for validation. Query and passage lengths are measured in words.}
    \label{tab:dataset_stats}
\end{table}
We conduct experiments on three diverse passage ranking datasets:
\begin{enumerate}
\item \textbf{ANTIQUE} \cite{Hashemi:antique:2019} is a non-factoid question answering benchmark based on the questions and answers of \textit{Yahoo! Webscope L6}. The questions were filtered to remove ones too short or duplicate. A resulting sample of question-answer pairs was then judged by crowd workers who assigned one of four relevance labels to each pair. All questions have well-formed correct grammar. For the evaluation we follow the authors' recommendation and treat the two higher labels as relevant and the lower two labels as irrelevant.

\item \textbf{InsuranceQA} \cite{feng2015applying} is a dataset from the insurance domain released in 2015. For this work we use the second version, which comes with a predefined train-, dev- and testset. The dev- and testset include for each question the relevant answers as well as a pool of $n \in \{100, 500, 1000, 1500\}$ irrelevant candidate answers. For our experiments, we choose $n = 500$. All queries and passages in this dataset consist of gramatically well-formed sentences.

\item \textbf{TREC-DL 2019} is the passage ranking dataset from the TREC deep learning track. It uses MS MARCO \cite{nguyen2016ms}, a collection of large \textit{Machine Reading Comprehension} datasets released by Microsoft in 2016.\footnote{\url{http://www.msmarco.org/}} This dataset was created using real, anonymized queries from the Bing search engine. The authors automatically identified queries that represented questions and extracted passages from the top-$10$ search results. These passages then manually received relevance labels from human annotators. The result is a very large dataset with over 8M passages and 1M queries. However, a number of queries have no associated relevant passages. Because of the nature of this dataset, queries and passages are not guaranteed to be grammatically or structurally correct or even made of complete sentences.
\end{enumerate}
Table \ref{tab:dataset_stats} outlines some dataset statistics. The evaluation (except for the ANTIQUE testset) follows the telescoping setting \cite{Matveeva06}, where a first round of retrieval has already been performed to select candidate passages that are relevant to the queries, followed by a re-ranking step by our models.

\section{Baselines}
Since we are mainly interested in improving the effectiveness and training efficiency of BERT-based models, the most important baseline is a vanilla \textbf{BERT} ranker \cite{nogueira2019passage}. The ranking is solely based on the output corresponding to the classification token, which is transformed into a scalar score using a feed-forward classification layer. Additionally, we implement other neural baselines:

\begin{enumerate}
\item \textbf{QA-LSTM} \cite{tan2016improved} is based on bidirectional LSTMs and attention. Both query and document are encoded using a shared bidirectional many-to-many LSTM and a pooling operation (maximum or average pooling) to the LSTM outputs. Attention scores are computed using the hidden LSTM states of the document and the pooled query representation. The resulting vectors are then compared using cosine similarity after applying dropout. We set the batch size to $32$ and the number of LSTM hidden units to $256$. We feed $300$-dimensional pre-trained GloVe \cite{pennington2014glove} embeddings to the shared LSTM and use a dropout rate of $0.5$.

\item \textbf{K-NRM} \cite{xiong2017end} is a neural ranking model that works via kernel pooling. Starting from pre-trained word embeddings, it builds a \textit{translation matrix}, where each row contains the cosine similarities of a query word to all document words. Each row is then fed into $K$ kernel functions, and the results are pooled by summation. Finally, a single transformation with tanh activation is applied to output a score. The model is trained with a pairwise ranking loss and uses RBF kernels. We use $300$-dimensional pre-trained GloVe embeddings to build the translation matrix. The hyperparameters are adopted from \cite{xiong2017end}: We set $K = 11$ and use one kernel for exact matches, i.e.\ $\mu_0 = 1$ and $\sigma_0 = 10^{-3}$. The remaining kernels are spaced evenly in $[-1, 1]$ with $\mu_1 = 0.9$, $\mu_2 = 0.7$, ..., $\mu_{10} = -0.9$ and $\sigma_1 = ... = \sigma_{10} = 0.1$. We use the Adam optimizer with a leaning rate of $0.001$ and $\epsilon = 10^{-5}$ and a batch size of $16$.

\item \textbf{Dynamic Memory Network} \cite{kumar2016ask,xiong2016dynamic} serves (in a slightly modified fashion) as the aggregation part of our model, which transforms sentence-level BERT outputs into a relevance score. We also train this model using pre-trained $300$-dimensional word vectors in order to analyze if and how much BERT representations improve the performance. For these experiments we use the same DMN hyperparameters as in our experiments with BERT-DMN to make the results more comparable.
\end{enumerate}

\subsection{Training Efficiency}
As previously mentioned, a drawback of BERT-based models is their training inefficiency, as the time required for even a single training epoch can be substantial, albeit a one-time cost. In order to mitigate this, we propose \bdlite{}. While the model architecture remains identical, the BERT layer is excluded from backpropagation, such that its weights remain frozen. This reduces the training time in two ways: The time required to complete the first epoch will be slightly lower, as the majority of the weights are excluded from the backward pass; the second and all subsequent epochs can be sped up significantly, as the BERT outputs can be cached and re-used.

\subsection{Training Details}
Our models are implemented using PyTorch. We use a pre-trained, uncased $\textup{BERT}_\textup{Base}$ model with $12$ encoder layers, $12$ attention heads and $768$-dimensional vector representations. The training is done as follows: We feed all query-passage pairs through the BERT layer to obtain the token representations. We then compute the average of all vectors for each sentence to obtain the inputs for the GRU, which in turn produces representations that serve as the inputs of the episodic memory module. Similarly, we use another GRU to encode the query as a single vector. In the case of \bdmn{}, the fine-tuning of \bert{} and training of the DMN happens jointly. For \bdlite{}, all weights corresponding to BERT are frozen, i.e.\ they remain unchanged during the optimization. BERT inputs are truncated if they exceed $512$ tokens.

The models are trained using the AdamW optimizer with the learning rate set to $3 \cdot 10^{-5}$, following \cite{nogueira2019passage}, and a pairwise max-margin loss: Let $q$ be a query and $p^+$ and $p^-$ passages, where $p^+$ is more relevant to $q$ than $p^-$. The loss is computed as
\begin{equation}
    \mathcal{L} = \max \left\{0, m - R \left(q, p^+ \right) + R \left(q, p^- \right) \right\}
\end{equation}
where $m$ is the margin and $R$ is the model. We use $m = 0.2$ and linear warm-up over the first $1000$ steps ($10000$ on TREC-DL). The DMN hyperparameters are set to $4$ episodes, $256$-dimensional hidden representations and a dropout rate of $0.1$. Dropout is applied at the DMN input, over the attention gates and before the output layer. We use a batch size of $32$ throughout our experiments. Validation is performed based on MAP on the devset. We use the same fixed random seed and thus identical training data for all experiments.

\subsection{Metrics}
The \textit{mean reciprocal rank} (MRR) is defined as
\begin{equation}
    \text{MRR} = \frac{1}{|Q|} \sum_{i=1}^{|Q|} \frac{1}{\text{rank}_i}
\end{equation}
where $Q$ is the set of all queries and $\text{rank}_i$ refers to the highest rank of any relevant document for the $i$-th query.

Similarly, \textit{mean average precision} (MAP) is defined as
\begin{equation}
    \text{AP}(q) = \frac{1}{|R_q|} \sum_{k=1}^{n} \text{P}(k) \times \text{rel}(k)
\end{equation}
\begin{equation}
    \text{MAP} = \frac{1}{|Q|} \sum_{q \in Q} \text{AP}(q)
\end{equation}
where $R_q$ is the set of all documents relevant to $q$, $n$ is the total number of retrieved documents, $\text{P}(k)$ is the precision and $\text{rel}(k)$ indicates the relevance of the document at rank $k$.

\section{Results}
In this section we present and discuss our results.

\subsection{Passage Re-Ranking Performance}
\begin{table}
    \centering
    \begin{tabular}{lccccccc}
        \toprule
                    & \multicolumn{2}{c}{ANTIQUE}
                    & \multicolumn{2}{c}{InsuranceQA}
                    & \multicolumn{2}{c}{TREC-DL} \\
                    \cmidrule(lr){2-3}
                    \cmidrule(lr){4-5}
                    \cmidrule(lr){6-7}
                    & MAP           & MRR               & MAP           & MRR           & MAP           & MRR \\
        \midrule
        QA-LSTM     & 0.488         & 0.619             & 0.185         & 0.231         & 0.193         & 0.519 \\
        K-NRM       & 0.511         & 0.654             & 0.176         & 0.215         & 0.237         & 0.567 \\
        DMN         & 0.491         & 0.613             & 0.092         & 0.118         & 0.136         & 0.274 \\
        \midrule
        \blite      & 0.593         & 0.774             & 0.259         & 0.314         & 0.327         & 0.739 \\
        \bdlite     & 0.675         & 0.851             & 0.374         & 0.449         & 0.418         & 0.859 \\
        \midrule
        \bert       & 0.697         & 0.849             & 0.399         & 0.476         & \bf 0.428     & 0.831 \\
        \bdmn       & \bf 0.700     & \bf 0.866         & \bf 0.406     & \bf 0.484     & 0.408         & \bf 0.889 \\
        \bottomrule
    \end{tabular}
    \caption{Passage re-ranking performance. \blite{} and \bdlite{} are architecturally identical to \bert{} and \bdmn{}, respectively, but only the classification layer is trained, while all other weights remain frozen. The baselines use pre-trained GloVe embeddings.}
    \label{tab:results}
\end{table}
Table \ref{tab:results} outlines the passage re-ranking performance of our methods and the baselines on three datasets. It is evident that the BERT-based methods vastly outperform the other baselines on all datasets. \blite{} performs noticeably worse, but still shows improvements over the non-contextual baselines. Finally, \bdmn{} improves the performance of \bert{} in all but one case. These results yield the following insights:
\begin{enumerate}
    \item As expected, the contextual token representations obtained from BERT trump non-contextual word embeddings. Even without any fine-tuning, the BERT representations perform well.
    \item The contextual sentence representations do in fact hold valuable information. This information is discarded by models which only use the output corresponding to the classification token. End-to-end training further improves the performance.
\end{enumerate}
As a result, the DMN profits vastly from BERT representations (\bdlite{}), and the performance improves even more when the model is trained end-to-end (\bdmn{}).

\subsection{The Effect of Fine-Tuning}
\begin{figure}
	\centering
     \includegraphics[width=0.7\linewidth]{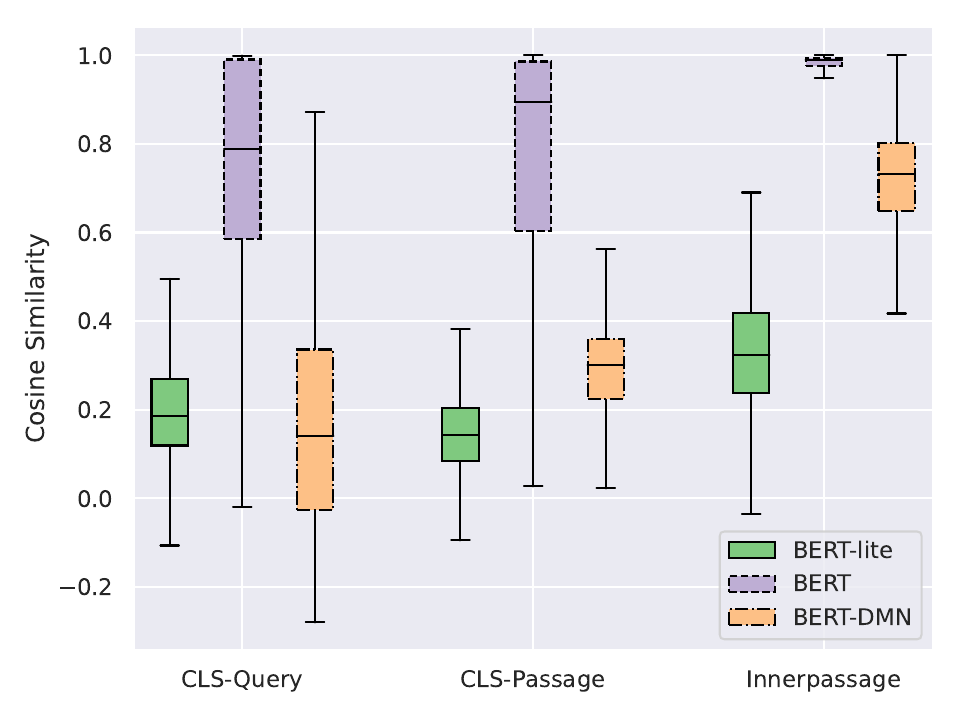}
	\caption{The diffusion of information within BERT representations on TREC-DL, illustrated by cosine similarities between classification token, query tokens and passage tokens.} 
	\label{fig:similarities}
\end{figure}
In order to analyze the effect of fine-tuning the parameters of \bert{}, we conduct additional experiments using \textit{lite} versions of \bert{} and \bdmn{}. The architectures and hyperparameters of these models are unchanged, however the number of trainable parameters is reduced roughly from 110M to 3M (\bdlite{}) or 1k (\blite{}) by freezing the BERT model itself. Table \ref{tab:results} shows slight performance drops of \bdlite{} in all but one case. However, comparing it to the fine-tuned vanilla \bert{} model shows even smaller differences, and in some cases the performance increases. Conversely, \blite{} exhibits a much higher loss of performance over \bert{}. This indicates that most of the information required for the task is already inherent to the pre-trained BERT model, and fine-tuning its parameters is merely required to direct it towards the desired output (usually the classification token). In order to confirm this hypothesis, we adopt a method proposed by \citet{pmlr-v119-goyal20a} to measure the \textit{diffusion of information} within the contextual representations output by BERT: Given a query-passage pair, we use a BERT model to obtain a representation (in our case a $768$-dimensional vector) of each token, corresponding to either query or passage. We then use cosine similarity to compute diffusion of information in three ways:
\begin{enumerate}
    \item \textbf{CLS-Query}: Cosine similarity between the classification token and each query token.
    \item \textbf{CLS-Passage}: Cosine similarity between the classification token and each passage token.
    \item \textbf{Innerpassage}: Cosine similarity between each possible pair of two passage tokens.
\end{enumerate}
The results are illustrated in Figure \ref{fig:similarities} for three BERT models, one without any fine-tuning (\blite{}), one with standard fine-tuning using only the classification output (\bert{}) and finally one fine-tuned as part of our approach (\bdmn{}). These measurements were performed on roughly $10\%$ of the TREC-DL testset (20k query-passage pairs). We observe that, without any fine-tuning, the outputs are rather dissimilar; with standard fine-tuning, however, the similarity of all representations vastly increases, especially within the passages. The same trend is exhibited by the model fine-tuned with \bdmn{}, but to a much lesser extend. This shows that discarding all but one output during fine-tuning leads to very high diffusion, in that all output vectors become very similar, and taking all outputs into account during fine-tuning alleviates this issue, allowing for a slight performance gain. It further suggests that \bdlite{} is able to combine the classification output and the sentence representations, performing closely to a fine-tuned BERT model.

\subsection{Training Efficiency}
\begin{table}
    \centering
    \begin{tabular}{lcc}
        \toprule
                & ANTIQUE                   & InsuranceQA \\
        \midrule
        \bert   & 1.71                      & 1.69 \\
        \bdlite & 2.26 $\rightarrow$ 5.32   & 2.55 $\rightarrow$ 5.67 \\
        \bottomrule
    \end{tabular}
    \caption{The average number of training batches (size 16) per second (higher is better). For \bdlite{}, we report one number for the first epoch and one number for all subsequent epochs.}
    \label{tab:efficiency}
\end{table}
Since the performances of \bdlite{} and \bert{} are comparable (cf. Table \ref{tab:results}), \bdlite{} can be seen as an alternative to the usual fine-tuning of a BERT model. Since the DMN layer has very few parameters compared to BERT (roughly 3M vs. 100M), the size of the model itself does not change a lot. However, \bdlite{} exhibits noticeable improvements in training efficiency compared to fine-tuning BERT. In order to show this, we measure the number of batches per seconds for both models in Table \ref{tab:efficiency}. For \bdlite{}, the first epoch is already slightly faster, as the majority of the weights are excluded from the backward pass; the second and all subsequent epochs are sped up significantly, as the BERT outputs can be cached re-used for the remainder of the training. The measurements were performed on a single non-shared NVIDIA GTX 1080Ti GPU.

\section{Conclusion and Outlook}
The exponential growth in the searchable web~\citep{holzmann2016dawn} has resulted in the proliferation of numerous knowledge-intensive tasks~\cite{holzmann2017exploring,singh2016expedition}, of which question answering tasks are prominent~\citep{nguyen2016ms_marco,anand2020conversational}. In this paper we introduced \bdmn{} and \bdlite{}, extensions of BERT that utilize dynamic memory networks to perform passage re-ranking. We have shown that our model improves the performance of BERT on three datasets. Moreover, \bdlite{} performs well even without a fine-tuned BERT model, reducing the training time while incurring only a small performance hit. Our findings demonstrate that fine-tuning BERT-based models is not always necessary, as nearly the same result can be achieved using sentence-level representations.

There are many ways to extend \bdmn{}. Firstly, a common problem of over-parameterized models like BERT is that they are less interpretable. There is some initial work in the direction of understanding the rationale behind QA and passage ranking tasks by either sparsification~\citep{zhang2021explain}, inspecting BERT's parametric memory~\citep{wallat2020bertnesia}, or in a post-hoc manner~\citep{zeon2019study,singh2020model:prefcov}. We see the DMN as an interpretable approach to evidence selection for question answering. The dynamic memory module in some sense iteratively computes attention on sentences that reflects their relative importance. We could use this observation to build an interpretable-by-design approach to passage ranking given questions by highlighting evidence sentences from the episodic memory module. Secondly, outside of text datasets, we envision the utility of the DMN in question answering over semi-structured data on the web like anchor text~\citep{holzmann2016tempas}, semantic annotations~\citep{holzmann2017exploring}, tables~\citep{fetahu2019tablenet} and fully structured knowledge graphs. Specifically, the transitive reasoning capability is natural to structured information organized as triples or in a graph.


\bibliography{references}

\begin{thebibliography}{51}
\expandafter\ifx\csname natexlab\endcsname\relax\def\natexlab#1{#1}\fi
\providecommand{\url}[1]{\texttt{#1}}
\providecommand{\href}[2]{#2}
\providecommand{\path}[1]{#1}
\providecommand{\DOIprefix}{doi:}
\providecommand{\ArXivprefix}{arXiv:}
\providecommand{\URLprefix}{URL: }
\providecommand{\Pubmedprefix}{pmid:}
\providecommand{\doi}[1]{\href{http://dx.doi.org/#1}{\path{#1}}}
\providecommand{\Pubmed}[1]{\href{pmid:#1}{\path{#1}}}
\providecommand{\bibinfo}[2]{#2}
\ifx\xfnm\relax \def\xfnm[#1]{\unskip,\space#1}\fi
\bibitem[{Devlin et~al.(2018)Devlin, Chang, Lee, and
  Toutanova}]{devlin2018bert}
\bibinfo{author}{J.~Devlin}, \bibinfo{author}{M.-W. Chang},
  \bibinfo{author}{K.~Lee}, \bibinfo{author}{K.~Toutanova},
\newblock \bibinfo{title}{Bert: Pre-training of deep bidirectional transformers
  for language understanding},
\newblock \bibinfo{journal}{arXiv preprint arXiv:1810.04805}
  (\bibinfo{year}{2018}).
\bibitem[{Kumar et~al.(2016)Kumar, Irsoy, Ondruska, Iyyer, Bradbury, Gulrajani,
  Zhong, Paulus, and Socher}]{kumar2016ask}
\bibinfo{author}{A.~Kumar}, \bibinfo{author}{O.~Irsoy},
  \bibinfo{author}{P.~Ondruska}, \bibinfo{author}{M.~Iyyer},
  \bibinfo{author}{J.~Bradbury}, \bibinfo{author}{I.~Gulrajani},
  \bibinfo{author}{V.~Zhong}, \bibinfo{author}{R.~Paulus},
  \bibinfo{author}{R.~Socher},
\newblock \bibinfo{title}{Ask me anything: Dynamic memory networks for natural
  language processing},
\newblock in: \bibinfo{booktitle}{International conference on machine
  learning}, \bibinfo{year}{2016}, pp. \bibinfo{pages}{1378--1387}.
\bibitem[{Xiong et~al.(2016)Xiong, Merity, and Socher}]{xiong2016dynamic}
\bibinfo{author}{C.~Xiong}, \bibinfo{author}{S.~Merity},
  \bibinfo{author}{R.~Socher},
\newblock \bibinfo{title}{Dynamic memory networks for visual and textual
  question answering},
\newblock in: \bibinfo{booktitle}{International conference on machine
  learning}, \bibinfo{year}{2016}, pp. \bibinfo{pages}{2397--2406}.
\bibitem[{Voorhees(1999)}]{Voorhees99thetrec-8}
\bibinfo{author}{E.~M. Voorhees},
\newblock \bibinfo{title}{The trec-8 question answering track report},
\newblock in: \bibinfo{booktitle}{In Proceedings of TREC-8},
  \bibinfo{year}{1999}, pp. \bibinfo{pages}{77--82}.
\bibitem[{Kwok et~al.(2001)Kwok, Etzioni, Etzioni, and Weld}]{kwok2001scaling}
\bibinfo{author}{C.~Kwok}, \bibinfo{author}{O.~Etzioni},
  \bibinfo{author}{O.~Etzioni}, \bibinfo{author}{D.~S. Weld},
\newblock \bibinfo{title}{Scaling question answering to the web},
\newblock \bibinfo{journal}{ACM Transactions on Information Systems (TOIS)}
  \bibinfo{volume}{19} (\bibinfo{year}{2001}) \bibinfo{pages}{242--262}.
\bibitem[{Tan et~al.(2016)Tan, Dos~Santos, Xiang, and Zhou}]{tan2016improved}
\bibinfo{author}{M.~Tan}, \bibinfo{author}{C.~Dos~Santos},
  \bibinfo{author}{B.~Xiang}, \bibinfo{author}{B.~Zhou},
\newblock \bibinfo{title}{Improved representation learning for question answer
  matching},
\newblock in: \bibinfo{booktitle}{Proceedings of the 54th Annual Meeting of the
  Association for Computational Linguistics (Volume 1: Long Papers)},
  \bibinfo{year}{2016}, pp. \bibinfo{pages}{464--473}.
\bibitem[{Tran and Nieder{\'e}e(2018)}]{tran2018multihop}
\bibinfo{author}{N.~K. Tran}, \bibinfo{author}{C.~Nieder{\'e}e},
\newblock \bibinfo{title}{Multihop attention networks for question answer
  matching},
\newblock in: \bibinfo{booktitle}{The 41st International ACM SIGIR Conference
  on Research \& Development in Information Retrieval},
  \bibinfo{organization}{ACM}, \bibinfo{year}{2018}, pp.
  \bibinfo{pages}{325--334}.
\bibitem[{Wang et~al.(2018{\natexlab{a}})Wang, Yu, Jiang, Zhang, Guo, Chang,
  Wang, Klinger, Tesauro, and Campbell}]{wang2018evidence}
\bibinfo{author}{S.~Wang}, \bibinfo{author}{M.~Yu}, \bibinfo{author}{J.~Jiang},
  \bibinfo{author}{W.~Zhang}, \bibinfo{author}{X.~Guo},
  \bibinfo{author}{S.~Chang}, \bibinfo{author}{Z.~Wang},
  \bibinfo{author}{T.~Klinger}, \bibinfo{author}{G.~Tesauro},
  \bibinfo{author}{M.~Campbell},
\newblock \bibinfo{title}{Evidence aggregation for answer re-ranking in
  open-domain question answering},
\newblock in: \bibinfo{booktitle}{International Conference on Learning
  Representations}, \bibinfo{year}{2018}{\natexlab{a}}.
\bibitem[{Wang et~al.(2018{\natexlab{b}})Wang, Yu, Guo, Wang, Klinger, Zhang,
  Chang, Tesauro, Zhou, and Jiang}]{wang2018r}
\bibinfo{author}{S.~Wang}, \bibinfo{author}{M.~Yu}, \bibinfo{author}{X.~Guo},
  \bibinfo{author}{Z.~Wang}, \bibinfo{author}{T.~Klinger},
  \bibinfo{author}{W.~Zhang}, \bibinfo{author}{S.~Chang},
  \bibinfo{author}{G.~Tesauro}, \bibinfo{author}{B.~Zhou},
  \bibinfo{author}{J.~Jiang},
\newblock \bibinfo{title}{R 3: Reinforced ranker-reader for open-domain
  question answering},
\newblock in: \bibinfo{booktitle}{Thirty-Second AAAI Conference on Artificial
  Intelligence}, \bibinfo{year}{2018}{\natexlab{b}}.
\bibitem[{Lin et~al.(2018)Lin, Ji, Liu, and Sun}]{lin2018denoising}
\bibinfo{author}{Y.~Lin}, \bibinfo{author}{H.~Ji}, \bibinfo{author}{Z.~Liu},
  \bibinfo{author}{M.~Sun},
\newblock \bibinfo{title}{Denoising distantly supervised open-domain question
  answering},
\newblock in: \bibinfo{booktitle}{Proceedings of the 56th Annual Meeting of the
  Association for Computational Linguistics (Volume 1: Long Papers)},
  \bibinfo{year}{2018}, pp. \bibinfo{pages}{1736--1745}.
\bibitem[{Chen et~al.(2017)Chen, Fisch, Weston, and Bordes}]{chen2017reading}
\bibinfo{author}{D.~Chen}, \bibinfo{author}{A.~Fisch},
  \bibinfo{author}{J.~Weston}, \bibinfo{author}{A.~Bordes},
\newblock \bibinfo{title}{Reading wikipedia to answer open-domain questions},
\newblock in: \bibinfo{booktitle}{Proceedings of the 55th Annual Meeting of the
  Association for Computational Linguistics (Volume 1: Long Papers)},
  \bibinfo{year}{2017}, pp. \bibinfo{pages}{1870--1879}.
\bibitem[{Wang et~al.(2017)Wang, Yuan, and Trischler}]{wang2017joint}
\bibinfo{author}{T.~Wang}, \bibinfo{author}{X.~Yuan},
  \bibinfo{author}{A.~Trischler},
\newblock \bibinfo{title}{A joint model for question answering and question
  generation},
\newblock \bibinfo{journal}{arXiv preprint arXiv:1706.01450}
  (\bibinfo{year}{2017}).
\bibitem[{Kratzwald and
  Feuerriegel(2018)}]{kratzwald-feuerriegel-2018-adaptive}
\bibinfo{author}{B.~Kratzwald}, \bibinfo{author}{S.~Feuerriegel},
\newblock \bibinfo{title}{Adaptive document retrieval for deep question
  answering},
\newblock in: \bibinfo{booktitle}{Proceedings of the 2018 Conference on
  Empirical Methods in Natural Language Processing},
  \bibinfo{publisher}{Association for Computational Linguistics},
  \bibinfo{address}{Brussels, Belgium}, \bibinfo{year}{2018}, pp.
  \bibinfo{pages}{576--581}. \URLprefix
  \url{https://www.aclweb.org/anthology/D18-1055}.
  \DOIprefix\doi{10.18653/v1/D18-1055}.
\bibitem[{Xu et~al.(2019)Xu, Ma, Nallapati, and Xiang}]{xu2019passage}
\bibinfo{author}{P.~Xu}, \bibinfo{author}{X.~Ma},
  \bibinfo{author}{R.~Nallapati}, \bibinfo{author}{B.~Xiang},
\newblock \bibinfo{title}{Passage ranking with weak supervision},
\newblock in: \bibinfo{booktitle}{International Conference on Learning
  Representations}, \bibinfo{year}{2019}.
\bibitem[{Guo et~al.(2020)Guo, Liu, Wang, Gao, Sankar, Yang, Guo, Zhang, Long,
  Chen, and Agarwal}]{guo2020detext}
\bibinfo{author}{W.~Guo}, \bibinfo{author}{X.~Liu}, \bibinfo{author}{S.~Wang},
  \bibinfo{author}{H.~Gao}, \bibinfo{author}{A.~Sankar},
  \bibinfo{author}{Z.~Yang}, \bibinfo{author}{Q.~Guo},
  \bibinfo{author}{L.~Zhang}, \bibinfo{author}{B.~Long}, \bibinfo{author}{B.-C.
  Chen}, \bibinfo{author}{D.~Agarwal},
\newblock \bibinfo{title}{Detext: A deep text ranking framework with bert},
\newblock in: \bibinfo{booktitle}{Proceedings of the 29th ACM International
  Conference on Information \& Knowledge Management}, CIKM '20,
  \bibinfo{publisher}{Association for Computing Machinery},
  \bibinfo{address}{New York, NY, USA}, \bibinfo{year}{2020}, p.
  \bibinfo{pages}{2509–2516}. \URLprefix
  \url{https://doi.org/10.1145/3340531.3412699}.
  \DOIprefix\doi{10.1145/3340531.3412699}.
\bibitem[{Zhan et~al.(2020)Zhan, Mao, Liu, Zhang, and Ma}]{zhan2020analysis}
\bibinfo{author}{J.~Zhan}, \bibinfo{author}{J.~Mao}, \bibinfo{author}{Y.~Liu},
  \bibinfo{author}{M.~Zhang}, \bibinfo{author}{S.~Ma},
\newblock \bibinfo{title}{An analysis of bert in document ranking},
\newblock in: \bibinfo{booktitle}{Proceedings of the 43rd International ACM
  SIGIR Conference on Research and Development in Information Retrieval}, SIGIR
  '20, \bibinfo{publisher}{Association for Computing Machinery},
  \bibinfo{address}{New York, NY, USA}, \bibinfo{year}{2020}, p.
  \bibinfo{pages}{1941–1944}. \URLprefix
  \url{https://doi.org/10.1145/3397271.3401325}.
  \DOIprefix\doi{10.1145/3397271.3401325}.
\bibitem[{Peters et~al.(2019)Peters, Ruder, and
  Smith}]{DBLP:conf/rep4nlp/PetersRS19}
\bibinfo{author}{M.~E. Peters}, \bibinfo{author}{S.~Ruder},
  \bibinfo{author}{N.~A. Smith},
\newblock \bibinfo{title}{To tune or not to tune? adapting pretrained
  representations to diverse tasks},
\newblock in: \bibinfo{editor}{I.~Augenstein}, \bibinfo{editor}{S.~Gella},
  \bibinfo{editor}{S.~Ruder}, \bibinfo{editor}{K.~Kann},
  \bibinfo{editor}{B.~Can}, \bibinfo{editor}{J.~Welbl},
  \bibinfo{editor}{A.~Conneau}, \bibinfo{editor}{X.~Ren},
  \bibinfo{editor}{M.~Rei} (Eds.), \bibinfo{booktitle}{Proceedings of the 4th
  Workshop on Representation Learning for NLP, RepL4NLP@ACL 2019, Florence,
  Italy, August 2, 2019.}, \bibinfo{publisher}{Association for Computational
  Linguistics}, \bibinfo{year}{2019}, pp. \bibinfo{pages}{7--14}. \URLprefix
  \url{https://www.aclweb.org/anthology/W19-4302/}.
\bibitem[{MacAvaney et~al.(2019)MacAvaney, Yates, Cohan, and
  Goharian}]{macavaney2019contextualized:bertir}
\bibinfo{author}{S.~MacAvaney}, \bibinfo{author}{A.~Yates},
  \bibinfo{author}{A.~Cohan}, \bibinfo{author}{N.~Goharian},
\newblock \bibinfo{title}{Contextualized word representations for document
  re-ranking},
\newblock \bibinfo{journal}{arXiv preprint arXiv:1904.07094}
  (\bibinfo{year}{2019}).
\bibitem[{Shen et~al.(2014)Shen, He, Gao, Deng, and Mesnil}]{Shen2014a}
\bibinfo{author}{Y.~Shen}, \bibinfo{author}{X.~He}, \bibinfo{author}{J.~Gao},
  \bibinfo{author}{L.~Deng}, \bibinfo{author}{G.~Mesnil},
\newblock \bibinfo{title}{A latent semantic model with convolutional-pooling
  structure for information retrieval},
\newblock in: \bibinfo{booktitle}{Proceedings of the 23rd ACM International
  Conference on Conference on Information and Knowledge Management}, CIKM '14,
  \bibinfo{publisher}{ACM}, \bibinfo{year}{2014}, pp.
  \bibinfo{pages}{101--110}. \URLprefix
  \url{http://doi.acm.org/10.1145/2661829.2661935}.
  \DOIprefix\doi{10.1145/2661829.2661935}.
\bibitem[{Huang et~al.(2013)Huang, He, Gao, Deng, Acero, and Heck}]{dssm13}
\bibinfo{author}{P.-S. Huang}, \bibinfo{author}{X.~He},
  \bibinfo{author}{J.~Gao}, \bibinfo{author}{L.~Deng},
  \bibinfo{author}{A.~Acero}, \bibinfo{author}{L.~Heck},
\newblock \bibinfo{title}{Learning deep structured semantic models for web
  search using clickthrough data},
\newblock in: \bibinfo{booktitle}{Proceedings of the 22nd ACM International
  Conference on Information and Knowledge Management}, CIKM '13,
  \bibinfo{publisher}{ACM}, \bibinfo{year}{2013}, pp.
  \bibinfo{pages}{2333--2338}. \URLprefix
  \url{http://doi.acm.org/10.1145/2505515.2505665}.
  \DOIprefix\doi{10.1145/2505515.2505665}.
\bibitem[{Shen et~al.(2014)Shen, He, Gao, Deng, and Mesnil}]{Shen2014b}
\bibinfo{author}{Y.~Shen}, \bibinfo{author}{X.~He}, \bibinfo{author}{J.~Gao},
  \bibinfo{author}{L.~Deng}, \bibinfo{author}{G.~Mesnil},
\newblock \bibinfo{title}{Learning semantic representations using convolutional
  neural networks for web search},
\newblock in: \bibinfo{booktitle}{Proceedings of the 23rd International
  Conference on World Wide Web}, WWW '14 Companion, \bibinfo{publisher}{ACM},
  \bibinfo{year}{2014}, pp. \bibinfo{pages}{373--374}. \URLprefix
  \url{http://doi.acm.org/10.1145/2567948.2577348}.
  \DOIprefix\doi{10.1145/2567948.2577348}.
\bibitem[{Xiong et~al.(2017)Xiong, Dai, Callan, Liu, and Power}]{KNRM17}
\bibinfo{author}{C.~Xiong}, \bibinfo{author}{Z.~Dai},
  \bibinfo{author}{J.~Callan}, \bibinfo{author}{Z.~Liu},
  \bibinfo{author}{R.~Power},
\newblock \bibinfo{title}{End-to-end neural ad-hoc ranking with kernel
  pooling},
\newblock in: \bibinfo{booktitle}{Proceedings of the 40th International ACM
  SIGIR Conference on Research and Development in Information Retrieval}, SIGIR
  '17, \bibinfo{publisher}{ACM}, \bibinfo{year}{2017}, pp.
  \bibinfo{pages}{55--64}. \URLprefix
  \url{http://doi.acm.org/10.1145/3077136.3080809}.
  \DOIprefix\doi{10.1145/3077136.3080809}.
\bibitem[{Guo et~al.(2016)Guo, Fan, Ai, and Croft}]{Guo2016}
\bibinfo{author}{J.~Guo}, \bibinfo{author}{Y.~Fan}, \bibinfo{author}{Q.~Ai},
  \bibinfo{author}{W.~B. Croft},
\newblock \bibinfo{title}{A deep relevance matching model for ad-hoc
  retrieval},
\newblock in: \bibinfo{booktitle}{Proceedings of the 25th ACM International on
  Conference on Information and Knowledge Management}, CIKM '16,
  \bibinfo{publisher}{ACM}, \bibinfo{year}{2016}, pp. \bibinfo{pages}{55--64}.
  \URLprefix \url{http://doi.acm.org/10.1145/2983323.2983769}.
  \DOIprefix\doi{10.1145/2983323.2983769}.
\bibitem[{Mitra et~al.(2017)Mitra, Diaz, and Craswell}]{mitra2017learning}
\bibinfo{author}{B.~Mitra}, \bibinfo{author}{F.~Diaz},
  \bibinfo{author}{N.~Craswell},
\newblock \bibinfo{title}{Learning to match using local and distributed
  representations of text for web search},
\newblock in: \bibinfo{booktitle}{Proceedings of the 26th International
  Conference on World Wide Web}, \bibinfo{organization}{International World
  Wide Web Conferences Steering Committee}, \bibinfo{year}{2017}, pp.
  \bibinfo{pages}{1291--1299}.
\bibitem[{Pang et~al.(2016)Pang, Lan, Guo, Xu, and Cheng}]{matchpyramid16}
\bibinfo{author}{L.~Pang}, \bibinfo{author}{Y.~Lan}, \bibinfo{author}{J.~Guo},
  \bibinfo{author}{J.~Xu}, \bibinfo{author}{X.~Cheng},
\newblock \bibinfo{title}{A study of {MatchPyramid} models on ad-hoc
  retrieval},
\newblock \bibinfo{journal}{SIGIR workshop on Neural Information Retrieval
  (NeuIR-16)} \bibinfo{volume}{arXiv:1606.04648} (\bibinfo{year}{2016}).
  \URLprefix \url{http://arxiv.org/abs/1606.04648}.
  \href{http://arxiv.org/abs/1606.04648}{{\tt arXiv:1606.04648}}.
\bibitem[{Nie et~al.(2018{\natexlab{a}})Nie, Li, and Nie}]{Nie_ictir18}
\bibinfo{author}{Y.~Nie}, \bibinfo{author}{Y.~Li}, \bibinfo{author}{J.-Y. Nie},
\newblock \bibinfo{title}{Empirical study of multi-level convolution models for
  ir based on representations and interactions},
\newblock in: \bibinfo{booktitle}{Proceedings of the 2018 ACM SIGIR
  International Conference on Theory of Information Retrieval}, ICTIR '18,
  \bibinfo{publisher}{ACM}, \bibinfo{year}{2018}{\natexlab{a}}, pp.
  \bibinfo{pages}{59--66}. \URLprefix
  \url{http://doi.acm.org/10.1145/3234944.3234954}.
  \DOIprefix\doi{10.1145/3234944.3234954}.
\bibitem[{Nie et~al.(2018{\natexlab{b}})Nie, Sordoni, and Nie}]{Nie_sigir_2018}
\bibinfo{author}{Y.~Nie}, \bibinfo{author}{A.~Sordoni}, \bibinfo{author}{J.-Y.
  Nie},
\newblock \bibinfo{title}{Multi-level abstraction convolutional model with weak
  supervision for information retrieval},
\newblock in: \bibinfo{booktitle}{the 41st International ACM SIGIR Conference},
  SIGIR '18, \bibinfo{publisher}{ACM}, \bibinfo{year}{2018}{\natexlab{b}}, pp.
  \bibinfo{pages}{985--988}. \URLprefix
  \url{http://doi.acm.org/10.1145/3209978.3210123}.
  \DOIprefix\doi{10.1145/3209978.3210123}.
\bibitem[{Hui et~al.(2017)Hui, Yates, Berberich, and de~Melo}]{pacrr17}
\bibinfo{author}{K.~Hui}, \bibinfo{author}{A.~Yates},
  \bibinfo{author}{K.~Berberich}, \bibinfo{author}{G.~de~Melo},
\newblock \bibinfo{title}{{PACRR}: A position-aware neural ir model for
  relevance matching},
\newblock in: \bibinfo{booktitle}{Proceedings of the 2017 Conference on
  Empirical Methods in Natural Language Processing},
  \bibinfo{address}{Copenhagen, Denmark}, \bibinfo{year}{2017}, pp.
  \bibinfo{pages}{1049--1058}. \URLprefix
  \url{https://www.aclweb.org/anthology/D17-1110}.
\bibitem[{Hui et~al.(2018)Hui, Yates, Berberich, and de~Melo}]{co_pacrr_wsdm18}
\bibinfo{author}{K.~Hui}, \bibinfo{author}{A.~Yates},
  \bibinfo{author}{K.~Berberich}, \bibinfo{author}{G.~de~Melo},
\newblock \bibinfo{title}{{Co-PACRR}: A context-aware neural ir model for
  ad-hoc retrieval},
\newblock in: \bibinfo{booktitle}{Proceedings of the Eleventh ACM International
  Conference on Web Search and Data Mining}, WSDM '18,
  \bibinfo{publisher}{ACM}, \bibinfo{year}{2018}, pp.
  \bibinfo{pages}{279--287}. \URLprefix
  \url{http://doi.acm.org/10.1145/3159652.3159689}.
  \DOIprefix\doi{10.1145/3159652.3159689}.
\bibitem[{McDonald et~al.(2018)McDonald, Brokos, and
  Androutsopoulos}]{pacrr_drmm_18}
\bibinfo{author}{R.~McDonald}, \bibinfo{author}{G.~Brokos},
  \bibinfo{author}{I.~Androutsopoulos},
\newblock \bibinfo{title}{Deep relevance ranking using enhanced document-query
  interactions},
\newblock in: \bibinfo{booktitle}{Proceedings of the 2018 Conference on
  Empirical Methods in Natural Language Processing}, \bibinfo{publisher}{ACL},
  \bibinfo{year}{2018}, pp. \bibinfo{pages}{1849--1860}. \URLprefix
  \url{http://aclweb.org/anthology/D18-1211}.
\bibitem[{Diaz et~al.(2016)Diaz, Mitra, and Craswell}]{diaz16}
\bibinfo{author}{F.~Diaz}, \bibinfo{author}{B.~Mitra},
  \bibinfo{author}{N.~Craswell},
\newblock \bibinfo{title}{Query expansion with locally-trained word
  embeddings},
\newblock in: \bibinfo{booktitle}{Proceedings of the 54th Annual Meeting of the
  Association for Computational Linguistics (Volume 1: Long Papers)},
  \bibinfo{year}{2016}, pp. \bibinfo{pages}{367--377}. \URLprefix
  \url{http://aclweb.org/anthology/P16-1035}.
  \DOIprefix\doi{10.18653/v1/P16-1035}.
\bibitem[{Zamani and Croft(2016)}]{Zamani_16a}
\bibinfo{author}{H.~Zamani}, \bibinfo{author}{W.~B. Croft},
\newblock \bibinfo{title}{Embedding-based query language models},
\newblock in: \bibinfo{booktitle}{Proceedings of the 2016 ACM International
  Conference on the Theory of Information Retrieval}, ICTIR '16,
  \bibinfo{publisher}{ACM}, \bibinfo{year}{2016}, pp.
  \bibinfo{pages}{147--156}. \URLprefix
  \url{http://doi.acm.org/10.1145/2970398.2970405}.
  \DOIprefix\doi{10.1145/2970398.2970405}.
\bibitem[{Hashemi et~al.(2019)Hashemi, Aliannejadi, Zamani, and
  Croft}]{Hashemi:antique:2019}
\bibinfo{author}{H.~Hashemi}, \bibinfo{author}{M.~Aliannejadi},
  \bibinfo{author}{H.~Zamani}, \bibinfo{author}{W.~B. Croft},
\newblock \bibinfo{title}{{ANTIQUE:} {A} non-factoid question answering
  benchmark},
\newblock \bibinfo{journal}{CoRR} \bibinfo{volume}{abs/1905.08957}
  (\bibinfo{year}{2019}). \URLprefix \url{http://arxiv.org/abs/1905.08957}.
\bibitem[{Feng et~al.(2015)Feng, Xiang, Glass, Wang, and
  Zhou}]{feng2015applying}
\bibinfo{author}{M.~Feng}, \bibinfo{author}{B.~Xiang}, \bibinfo{author}{M.~R.
  Glass}, \bibinfo{author}{L.~Wang}, \bibinfo{author}{B.~Zhou},
\newblock \bibinfo{title}{Applying deep learning to answer selection: A study
  and an open task},
\newblock in: \bibinfo{booktitle}{2015 IEEE Workshop on Automatic Speech
  Recognition and Understanding (ASRU)}, \bibinfo{organization}{IEEE},
  \bibinfo{year}{2015}, pp. \bibinfo{pages}{813--820}.
\bibitem[{Nguyen et~al.(2016)Nguyen, Rosenberg, Song, Gao, Tiwary, Majumder,
  and Deng}]{nguyen2016ms}
\bibinfo{author}{T.~Nguyen}, \bibinfo{author}{M.~Rosenberg},
  \bibinfo{author}{X.~Song}, \bibinfo{author}{J.~Gao},
  \bibinfo{author}{S.~Tiwary}, \bibinfo{author}{R.~Majumder},
  \bibinfo{author}{L.~Deng},
\newblock \bibinfo{title}{Ms marco: A human-generated machine reading
  comprehension dataset}  (\bibinfo{year}{2016}).
\bibitem[{Matveeva et~al.(2006)Matveeva, Burges, Burkard, Laucius, and
  Wong}]{Matveeva06}
\bibinfo{author}{I.~Matveeva}, \bibinfo{author}{C.~Burges},
  \bibinfo{author}{T.~Burkard}, \bibinfo{author}{A.~Laucius},
  \bibinfo{author}{L.~Wong},
\newblock \bibinfo{title}{High accuracy retrieval with multiple nested ranker},
\newblock in: \bibinfo{booktitle}{Proceedings of the 29th International ACM
  SIGIR Conference on Research and Development in Information Retrieval}, SIGIR
  '06, \bibinfo{publisher}{ACM}, \bibinfo{year}{2006}, pp.
  \bibinfo{pages}{437--444}. \URLprefix
  \url{http://doi.acm.org/10.1145/1148170.1148246}.
  \DOIprefix\doi{10.1145/1148170.1148246}.
\bibitem[{Nogueira and Cho(2019)}]{nogueira2019passage}
\bibinfo{author}{R.~Nogueira}, \bibinfo{author}{K.~Cho},
\newblock \bibinfo{title}{Passage re-ranking with bert},
\newblock \bibinfo{journal}{arXiv preprint arXiv:1901.04085}
  (\bibinfo{year}{2019}).
\bibitem[{Pennington et~al.(2014)Pennington, Socher, and
  Manning}]{pennington2014glove}
\bibinfo{author}{J.~Pennington}, \bibinfo{author}{R.~Socher},
  \bibinfo{author}{C.~D. Manning},
\newblock \bibinfo{title}{Glove: Global vectors for word representation},
\newblock in: \bibinfo{booktitle}{Empirical Methods in Natural Language
  Processing (EMNLP)}, \bibinfo{year}{2014}, pp. \bibinfo{pages}{1532--1543}.
  \URLprefix \url{http://www.aclweb.org/anthology/D14-1162}.
\bibitem[{Xiong et~al.(2017)Xiong, Dai, Callan, Liu, and Power}]{xiong2017end}
\bibinfo{author}{C.~Xiong}, \bibinfo{author}{Z.~Dai},
  \bibinfo{author}{J.~Callan}, \bibinfo{author}{Z.~Liu},
  \bibinfo{author}{R.~Power},
\newblock \bibinfo{title}{End-to-end neural ad-hoc ranking with kernel
  pooling},
\newblock in: \bibinfo{booktitle}{Proceedings of the 40th International ACM
  SIGIR conference on research and development in information retrieval},
  \bibinfo{organization}{ACM}, \bibinfo{year}{2017}, pp.
  \bibinfo{pages}{55--64}.
\bibitem[{Goyal et~al.(2020)Goyal, Choudhury, Raje, Chakaravarthy, Sabharwal,
  and Verma}]{pmlr-v119-goyal20a}
\bibinfo{author}{S.~Goyal}, \bibinfo{author}{A.~R. Choudhury},
  \bibinfo{author}{S.~Raje}, \bibinfo{author}{V.~Chakaravarthy},
  \bibinfo{author}{Y.~Sabharwal}, \bibinfo{author}{A.~Verma},
\newblock \bibinfo{title}{{P}o{WER}-{BERT}: Accelerating {BERT} inference via
  progressive word-vector elimination},
\newblock in: \bibinfo{editor}{H.~D. III}, \bibinfo{editor}{A.~Singh} (Eds.),
  \bibinfo{booktitle}{Proceedings of the 37th International Conference on
  Machine Learning}, volume \bibinfo{volume}{119} of
  \textit{\bibinfo{series}{Proceedings of Machine Learning Research}},
  \bibinfo{publisher}{PMLR}, \bibinfo{year}{2020}, pp.
  \bibinfo{pages}{3690--3699}. \URLprefix
  \url{http://proceedings.mlr.press/v119/goyal20a.html}.
\bibitem[{Holzmann et~al.(2016)Holzmann, Nejdl, and Anand}]{holzmann2016dawn}
\bibinfo{author}{H.~Holzmann}, \bibinfo{author}{W.~Nejdl},
  \bibinfo{author}{A.~Anand},
\newblock \bibinfo{title}{The dawn of today's popular domains: A study of the
  archived german web over 18 years},
\newblock in: \bibinfo{booktitle}{2016 IEEE/ACM Joint Conference on Digital
  Libraries (JCDL)}, \bibinfo{organization}{IEEE}, \bibinfo{year}{2016}, pp.
  \bibinfo{pages}{73--82}.
\bibitem[{Holzmann et~al.(2017)Holzmann, Nejdl, and
  Anand}]{holzmann2017exploring}
\bibinfo{author}{H.~Holzmann}, \bibinfo{author}{W.~Nejdl},
  \bibinfo{author}{A.~Anand},
\newblock \bibinfo{title}{Exploring web archives through temporal anchor
  texts},
\newblock in: \bibinfo{booktitle}{Proceedings of the 2017 ACM on Web Science
  Conference}, \bibinfo{year}{2017}, pp. \bibinfo{pages}{289--298}.
\bibitem[{Singh et~al.(2016)Singh, Nejdl, and Anand}]{singh2016expedition}
\bibinfo{author}{J.~Singh}, \bibinfo{author}{W.~Nejdl},
  \bibinfo{author}{A.~Anand},
\newblock \bibinfo{title}{Expedition: a time-aware exploratory search system
  designed for scholars},
\newblock in: \bibinfo{booktitle}{Proceedings of the 39th International ACM
  SIGIR conference on Research and Development in Information Retrieval},
  \bibinfo{year}{2016}, pp. \bibinfo{pages}{1105--1108}.
\bibitem[{Nguyen et~al.(2016)Nguyen, Rosenberg, Song, Gao, Tiwary, Majumder,
  and Deng}]{nguyen2016ms_marco}
\bibinfo{author}{T.~Nguyen}, \bibinfo{author}{M.~Rosenberg},
  \bibinfo{author}{X.~Song}, \bibinfo{author}{J.~Gao},
  \bibinfo{author}{S.~Tiwary}, \bibinfo{author}{R.~Majumder},
  \bibinfo{author}{L.~Deng},
\newblock \bibinfo{title}{{MS MARCO}: A human generated machine reading
  comprehension dataset},
\newblock \bibinfo{journal}{arXiv preprint arXiv:1611.09268}
  (\bibinfo{year}{2016}).
\bibitem[{Anand et~al.(2020)Anand, Cavedon, Joho, Sanderson, and
  Stein}]{anand2020conversational}
\bibinfo{author}{A.~Anand}, \bibinfo{author}{L.~Cavedon},
  \bibinfo{author}{H.~Joho}, \bibinfo{author}{M.~Sanderson},
  \bibinfo{author}{B.~Stein},
\newblock \bibinfo{title}{Conversational search (dagstuhl seminar 19461)},
\newblock in: \bibinfo{booktitle}{Dagstuhl Reports},
  volume~\bibinfo{volume}{9}, \bibinfo{organization}{Schloss
  Dagstuhl-Leibniz-Zentrum f{\"u}r Informatik}, \bibinfo{year}{2020}.
\bibitem[{Zhang et~al.(2021)Zhang, Rudra, and Anand}]{zhang2021explain}
\bibinfo{author}{Z.~Zhang}, \bibinfo{author}{K.~Rudra},
  \bibinfo{author}{A.~Anand},
\newblock \bibinfo{title}{Explain and predict, and then predict again},
\newblock in: \bibinfo{booktitle}{{WSDM} '21, The Fourteenth {ACM}
  International Conference on Web Search and Data Mining, Virtual Event,
  Israel, March 8-12, 2021}, \bibinfo{publisher}{{ACM}}, \bibinfo{year}{2021},
  pp. \bibinfo{pages}{418--426}. \URLprefix
  \url{https://doi.org/10.1145/3437963.3441758}.
  \DOIprefix\doi{10.1145/3437963.3441758}.
\bibitem[{Singh et~al.(2020)Singh, Wallat, and Anand}]{wallat2020bertnesia}
\bibinfo{author}{J.~Singh}, \bibinfo{author}{J.~Wallat},
  \bibinfo{author}{A.~Anand},
\newblock \bibinfo{title}{Bertnesia: Investigating the capture and forgetting
  of knowledge in bert},
\newblock in: \bibinfo{booktitle}{Proceedings of the Third BlackboxNLP Workshop
  on Analyzing and Interpreting Neural Networks for NLP}, \bibinfo{year}{2020},
  pp. \bibinfo{pages}{174--183}.
\bibitem[{Fernando et~al.(2019)Fernando, Singh, and Anand}]{zeon2019study}
\bibinfo{author}{Z.~T. Fernando}, \bibinfo{author}{J.~Singh},
  \bibinfo{author}{A.~Anand},
\newblock \bibinfo{title}{A study on the interpretability of neural retrieval
  models using deepshap},
\newblock in: \bibinfo{booktitle}{Proceedings of the 42nd International ACM
  SIGIR Conference on Research and Development in Information Retrieval},
  \bibinfo{year}{2019}, pp. \bibinfo{pages}{1005--1008}.
\bibitem[{Singh and Anand(2020)}]{singh2020model:prefcov}
\bibinfo{author}{J.~Singh}, \bibinfo{author}{A.~Anand},
\newblock \bibinfo{title}{Model agnostic interpretability of rankers via intent
  modelling},
\newblock in: \bibinfo{booktitle}{Proceedings of the 2020 Conference on
  Fairness, Accountability, and Transparency}, \bibinfo{year}{2020}, pp.
  \bibinfo{pages}{618--628}.
\bibitem[{Holzmann and Anand(2016)}]{holzmann2016tempas}
\bibinfo{author}{H.~Holzmann}, \bibinfo{author}{A.~Anand},
\newblock \bibinfo{title}{Tempas: Temporal archive search based on tags},
\newblock in: \bibinfo{booktitle}{Proceedings of the 25th International
  Conference Companion on World Wide Web}, \bibinfo{year}{2016}, pp.
  \bibinfo{pages}{207--210}.
\bibitem[{Fetahu et~al.(2019)Fetahu, Anand, and Koutraki}]{fetahu2019tablenet}
\bibinfo{author}{B.~Fetahu}, \bibinfo{author}{A.~Anand},
  \bibinfo{author}{M.~Koutraki},
\newblock \bibinfo{title}{Tablenet: An approach for determining fine-grained
  relations for wikipedia tables},
\newblock in: \bibinfo{booktitle}{The World Wide Web Conference},
  \bibinfo{year}{2019}, pp. \bibinfo{pages}{2736--2742}.

\end{thebibliography}


\end{document}